\documentclass[aps, prl, floatfix, preprint, showpacs]{revtex4-1}

\usepackage{graphicx}
\usepackage{color}
\usepackage{calc}
\usepackage{array}
\usepackage{graphicx}
\usepackage{amsmath, amssymb}
\usepackage{natbib}
\usepackage{hyperref}
\begin{document}
%%\special{header=psfrag.pro}
%% ==================================================================
%%\begin{titlepage}
%%
%%
\title{The value of curl(curl $A$) - grad(div $A$) + div(grad $A$) for an 
absolute vector $A$ }

\author{W.L.Kennedy}
\affiliation{Department of Physics,
             University of Otago,
             Dunedin, New Zealand}
%% ==================================================================
\begin{abstract}
The well-known identity involving the expression presented in the above title is considered in Riemannian 
and in Euclidean space without restriction on the coordinate system adopted therein. The Riemann and Ricci 
tensors intrinsically assume a defining role in the analysis.
The analysis is designed to put an end to the myriad of confusing and mostly incorrect statements about 
the identity, which are found in textbooks and in the literature. \end{abstract}

\date{\today}
%%
%%\end{titlepage}
\maketitle
{\bf{Introduction.}}

Take the oft-maligned so-called vector identity
\begin{equation}\label{EQ01}
( curl\ curl - grad\ div + div\ grad\ )\bf{A} = 0
\end{equation}
and denote its left-hand side by ${\bf{\Omega}\bf{A}}$.

Sometimes it is stated that the identity must be defined to be true
and in other places it is claimed that its truth can be defined only in a Cartesian system.  Often it is just the 
'div grad' term that is 'punished' by claims that it has meaning only in Cartesian coordinates and otherwise must be 
defined by Equation (\ref{EQ01}).{\cite{bib-name1}}, {\cite{bib-name2}}. We therefore propose to examine 
the identity in a general curvilinear coordinate system,(not necessarily orthogonal) for which we assume that 
there are no mathematical difficulties of note. 
Some supposed difficulties of a coordinate system might include for example: the singular point at the origin for a 
spherical polar system, the $z$-axis singular line for a cylindrical polar system, while spheroidal coordinate systems 
also have singularities.  
Such singularities however far from being an embarrassment in fact provide pointers towards the limiting
cases of systems for which a particular choice of coordinates may be useful:  For example spherical polar coordinates 
for point sources at the origin, cylindrical polar coordinates for infinite line sources along the symmetry axis,
while spheroidal coordinates also have useful singularities:  
The prolate spheroidal system has a finite singular line (a degenerate prolate ellipsoid) on the symmetry 
axis which suits boundary value problems involving a finite line source held at constant potential, 
while for the oblate system the ($x,y$) plane singular disc (a degenerate oblate ellipsoid) suits problems 
involving a disc source held at constant potential.
 
However before discussing the identity illustrated by Equation (\ref{EQ01}), in curvilinear coordinates in a 
Euclidean space, we first consider the more general question of the truth or otherwise of the identity in 
a Riemannian space $V$. In a Riemannian space we are forced to consider the proper tensor character of the 
various elements occurring in Equation (\ref{EQ01}), in a more general way than that usually adopted in 
a Euclidean space. Furthermore many Euclidean space treatments specialise by assuming an 
orthogonal curvilinear coordinate system instead of leaving the metric tensor unrestricted.
%%%%%%%%%%%%%%%%%%%%%%%%%%%%%%%%%%%%%%%%%%%%%%%%%%%%%%%%%%%%%%%%%%%%%%%%%%%%%%%%%%%%%%%%%%%%%%%%%%
%%%%%%%%%%%%%%%%%%%%%%%%%%%%%%%%%%%%%%%%%%%%%%%%%%%%%%%%%%%%%%%%%%%%%%%%%%%%%%%%%%%%%%%%%%%%%%%%%%

(a)\ \ \ {\bf{The Riemannian space discussion.}} \ \ \ 
%%%%%%%%%%%%%%%%%%%%%%%%%%%%%%%%%%%%%%%%%%%%%%%%%%%%%%%%%%%%%%%%%%%%%%%%%%%%%%%%%%%%%%%%%%%%%%%%5
The basic premise of any discussion of Equation (\ref{EQ01}), must be that each term should behave in the same way 
under coordinate transformations.
First consider the `grad-div' term. If $A$ is an absolute vector then requiring $div A$ to be an absolute 
scalar means that the differentiation involved with the divergence must be a covariant derivative.  The 
resulting scalar produces an absolute vector under the gradient operation with either covariant or regular 
partial derivatives being used. Now consider the `div-grad' term, the term which is most disputed in naive 
discussions of the identity, Equation (\ref{EQ01}). In order to produce an absolute 
vector from an absolute vector, the `div-grad' operation must be a scalar operation; in tensor analysis we 
must have this operation to be the contracted sequence of two covariant differentiations. Two independent 
covariant differentiations will produce an absolute rank 3 tensor; the desired contraction on the two 
differentiations then gives an absolute vector. 
From a tensor analysis point of view there is absolutely no problem with this set of operations. On the contrary 
it is the `curl-curl' term to which most attention must be paid.
The curl of an antisymmetric tensor of rank $m$ in an n--dimensional $V_n$ is a tensor of 
rank ($m$+1).  Since 'physics-vector-analysis' requires the curl of a vector to 
produce a vector, then the `physics curl' is to be found as the dual of the antisymmetric Stokes tensor, 
i.e. the antisymmetrized covariant derivative of the vector. Thus at this stage of our analysis any Riemannian $V$ 
that we consider can only be a Riemannian $V_3$. In terms of this physics `curl' we 
put  \begin{math}{\bf{B}} = curl\ {\bf{A}} \end{math} using
\begin{equation} \label{EQ02} B^\alpha =   (g^{-1/2})\ \varepsilon^{\alpha\beta\gamma}\ \nabla_
\beta A _\gamma   =   (g^{-1/2})\ \varepsilon^{\alpha\beta\gamma}\ \partial_
\beta A_\gamma \end{equation}
where the epsilon symbol $\varepsilon^{\alpha\beta\gamma}$  is the contravariant version 
of the Levi-Civita alternating tensor, $g$ is the determinant of the covariant metric tensor and the 
usual summation convention is assumed. 
(If we were to consider transformations between coordinate 
systems, the only restriction we would make is that `handedness' doesn't change so that the standard 
values for the elements of the epsilon tensor are unaffected.) The factor involving $g$ compensates 
for the weight, +1, of the contravariant epsilon tensor. since $g$ is a relative scalar of weight, +2. 
Thus from (\ref{EQ02}) an absolute vector $A$ produces the absolute vector $B$. 
The Christoffel symbol term involved in writing out the covariant derivative gives zero from 
symmetry/antisymmetry considerations so that the form with ordinary partial derivatives is also valid. Since 
two curl operations are involved in the first term of Equation (\ref{EQ01}), it is convenient to retain 
the covariant derivative 
form since we know that a covariant derivative operator may be freely moved to the left or right past any 
metric tensor factor or its determinant, $g$, or any function of $g$. Although the covariant derivative of 
the Levi-Civita tensor density is zero, this fact is not often remarked upon in tensor analysis treatments. 
Since the proof of this statement is a little messy due to the special values of the epsilon's components 
we adopt here the more economical device of establishing that the contracted covariant derivative 
of the epsilon tensor is zero. We thereby achieve sufficient freedom with respect to ordering and re--ordering 
the elements of Equation (\ref{EQ02}), for our purposes. The one-line proof is:
\begin{equation} \label{EQ03}
\nabla_\mu \varepsilon^{\lambda\mu\nu} = \partial_\mu \varepsilon ^{\lambda\mu\nu} - 
\Gamma^\alpha_{\mu \alpha }\varepsilon ^{\lambda\mu\nu} +
\Gamma^\lambda_{\mu \alpha }\varepsilon ^{\alpha\mu\nu} +
\Gamma^\mu_{\mu \alpha }\varepsilon ^{\lambda\alpha\nu} +
\Gamma^\nu_{\mu \alpha }\varepsilon ^{\lambda\mu\alpha} = 0 \end{equation}
In the expanded expression, the second and fourth terms cancel, and each of the other terms is separately 
zero. Note that the second term on the RHS of (\ref{EQ03}) is due to the weight of the epsilon tensor, and 
also that the choice of index pair for the contraction on the LHS is immaterial.
Now define  ${\bf{J}} = curl\ {\bf{B}} = curl\ (curl\ {\bf{A})}$ in the same way and in the same form 
on covariant components of $\bf{B}$ and use Equation (\ref{EQ03}) to shift just one covariant derivative 
operator fully to the left. Thus
\begin{equation} \label{EQ04}
J^\lambda = \nabla_\mu \ (( 1/g)\ \varepsilon^{\lambda\mu\nu} 
 g _{\nu \alpha}\ \varepsilon^{\alpha\beta\gamma} \nabla_\beta
 A _\gamma) \end{equation}
This expression is greatly simplified if we force $\bf{A}$ to occur as contravariant components; bring one 
set of epsilon indices down using $1/g$ as the determinant of the matrix of the 
contravariant $g^{\alpha \beta}$, namely,
\begin{equation} \label{EQ05} (1/g)\ \varepsilon^{\lambda\mu\nu}
=\varepsilon_{\xi\eta\zeta}\ g^{\lambda\xi} g^{\mu\eta} g^{\nu\zeta} \end{equation}
and then get
\begin{equation} \label{EQ06} J^\lambda = \nabla_\mu  ( \varepsilon^{\alpha\beta\gamma} 
  \varepsilon_{\xi\eta\zeta} 
  g^{\lambda \xi} g^{\mu \eta} g^{\nu \zeta}  g_{\nu \alpha} g_{\gamma \rho} \nabla_\beta A ^\rho) 
  =  \nabla_{\mu}\nabla^{\lambda}  A^{\mu}  -  \nabla_{\mu}  \nabla^{\mu} A^{\lambda}  \end{equation}
This gives us the `\emph{curl\ curl}' term of ${\bf{\Omega}}{\bf{A}}$.  We now add to this result the following 
contributions from the `\emph{grad\ div}' and `\emph{div\ grad}' terms,
\begin{equation} \label{EQ07} (-\mbox{grad div} {\bf{A}} + \mbox{div grad} {\bf{A}})^{\lambda}
                 =-\nabla^{\lambda}\nabla_{\mu} A^{\mu}+ 
                   \nabla_{\mu} \nabla^{\mu} A^{\lambda}  \end{equation}
and get
\begin{equation} \label{EQ08}
({\bf{\Omega A}})_\beta = g_{\beta \lambda} ({\bf{\Omega A}})^\lambda
   = g^{\sigma\mu} ( \nabla_{\mu} \nabla_{\beta} - \nabla_{\beta} \nabla_{\mu} ) A_{\sigma} =
   -g^{\sigma \mu} A^{\rho} R_{\sigma\rho\beta\mu} = -A^{\rho} R_{\rho\beta} \end{equation}
The rank-four Riemann curvature tensor $R$ in fully covariant form, automatically appears, by definition, 
from the double---covariant---derivative---commutator acting on the covariant vector ${A_{\sigma}}$. The 
contraction implicit in the fourth member of (\ref{EQ08}) produces the rank-two covariant Ricci 
tensor $R$.{\cite{bib-name3}}. It is historically conventional that these two tensors are symbolically 
distinguished only by their ranks. The equation, (\ref{EQ08}), can also be written as
\begin{equation} \label{EQ09}
({\bf{\Omega A}})_\beta = 
    ( \nabla^{\sigma} \nabla_{\beta} - \nabla_{\beta} \nabla^{\sigma} ) A_{\sigma} =
-A^{\rho} R_{\rho\beta} \end{equation}
showing the double--covariant derivative commutator acting in mixed form to directly produce the Ricci 
tensor form due to the contraction implicit in Equation (\ref{EQ09})
Thus in a genuinely Riemannian $V_3$ the value of the expression $\bf{\Omega}\bf{A}$, which is written out in the 
title of this paper is not zero {\cite{bib-name4}}, {\cite{bib-name5}}. Our conclusion is thus:
Equation (\ref{EQ01}) is not true in Riemannian space $V_3$.
Equations (\ref{EQ08}), (\ref{EQ09}) show that 
the space needs to be Ricci-flat for the identity, ${\bf{\Omega}\bf{A}} = 0$, to hold. Consider now the two 
concepts of Ricci--flatness (the Ricci tensor identically zero), and Riemann--flatness (the 
Riemann tensor identically zero) for a space $V$. 
In Einstein's general relativity which operates in pseudo-Riemannian space--time, regions of space--time 
which are source--free in the sense that there the energy-momentum tensor is zero, are Ricci--flat but 
not necessarily Riemann--flat. Space--time for such a region is allowed to be Ricci--flat while not 
being Riemann--flat, reflecting the effect of distant sources on a vacuum region. Generally for a 
Riemannian space $V_n$, with $n\geq4$, Ricci--flatness is not equivalent to Riemannian--flatness; however 
for a $V_3$ the number of independent Riemann components, six, equals the number of independent Ricci components 
and Ricci--flatness becomes equivalent to Riemann--flatness. This is easily proven:

Firstly, Riemann--flatness obviously generally 
implies Ricci--flatness; secondly, to confirm the converse, we just need to be able to invert the 
6$\times$6 matrix expressing the linear relation between independent Ricci components and independent Riemann 
components. The magnitude of the determinant of this matrix is easily found to be ${\pm}2/g^2$, where the 
algebraic sign depends on the ordering chosen for the Ricci and Riemann components in the array 
of connecting equations. The determinant of the matrix for the connecting equations being 
non-zero, our statement of equivalence is established. Thus $\bf{\Omega}\bf{A}$ = 0 also requires Riemann--flatness 
and the space has to be Euclidean.  We now consider Equation (\ref{EQ01}) directly in a purely Euclidean space.
  
%%%%%%%%%%%%%%%%%%%%%%%%%%%%%%%%%%%%%%%%%%%%%%%%%%%%%%%%%%%%%%%%%%%%%%%%%%%%%%%%%%%%%%%%%%%%%%%%%%%%%%%
%%%%%%%%%%%%%%%%%%%%%%%%%%%%%%%%%%%%%%%%%%%%%%%%%%%%%%%%%%%%%%%%%%%%%%%%%%%%%%%%%%%%%%%%%%%%%%%%%%%%%%%
(b): \ \ \ {\bf{The Euclidean space discussion.}} \ \ \ 
%%%%%%%%%%%%%%%%%%%%%%%%%%%%%%%%%%%%%%%%%%%%%%%%%%%%%%%%%%%%%%%%%%%%%%%%%%%%%%%%%%%%%%%%%%%%%%%%%%%%%%
Here we briefly indicate how in $E_3$, using a not necessarily orthogonal curvilinear coordinate system 
more accessible proofs are available but still without any necessity to explicitly calculate individually 
any of the three terms of ${\bf{\Omega}}\bf{A}$.  In $E_3$ one has the option of introducing a covariant basis 
vector set \{${\bf{e}_\alpha}$\} defined in the usual way 
via ${{\bf{e}}_\alpha = \partial{\bf{r}}/\partial x^\alpha}$. Modern usage would prefer changing this definition to 
 ${{\bf{e}}_\alpha = \partial/\partial x^\alpha}$ 
but the older style definition has the physical advantage of being able to explicitly show the basis vectors 
to scale on a 3D perspective sketch of the coordinate system. The metric tensor is then
\begin{equation} \label{EQ10} g_{\alpha\beta} = {\bf{e}}_\alpha . {\bf{e}}_\beta \end{equation}
The contravariant components $g^{\alpha\beta}$ of the tensor $\bf{g}$ have a matrix which is the inverse of the 
matrix of the ${g_{\alpha\beta}}$. The vectors of the contravariant basis set \{${\bf{e}^\alpha}$\} satisfy 
\begin{equation} \label{EQ11} {\bf{e}}^\alpha.{\bf{e}}_\beta =  \delta^\alpha_\beta
\end{equation}
and can be found using  
\begin{equation} \label{EQ12}  {\bf{e}}^\alpha = g^{\alpha\beta}{\bf{e}}_\beta  \end{equation}
However in $E_3$ we note that one may also calculate contravariant basis vectors by

\begin{equation} \label{EQ13} {\bf{e}}^\mu = (1/E)\ 
\varepsilon^{\mu\alpha\beta} {\bf{e}}_\alpha\times{\bf{e}}_\beta \end{equation}
where $E$ is the scalar triple product of the covariant basis set,
\begin{equation} \label{EQ14} [{\bf{e}}_1,{\bf{e}}_2,{\bf{e}}_3]\equiv E = \surd{g}
\end{equation}
The gradient operator can be written as either ${\bf{e}}^\alpha\partial_\alpha$, or 
as ${\bf{e}}_\alpha\partial^\alpha$. We can push a $\partial$ past a base vector 
using the commutators
\begin{equation} \label{EQ15}
\partial_\alpha{\bf{e}}^\beta - {\bf{e}}^\beta   \partial_\alpha
=-\Gamma^\beta_{\alpha\lambda}{\bf e}^\lambda \end{equation}
\begin{equation} \label{EQ16}
\partial_\alpha{\bf{e}}_\beta - {\bf{e}}_\beta  \partial_\alpha
= \Gamma^\lambda_{\alpha\beta}{\bf e}_\lambda \end{equation}
and thus see that
\begin{equation} \label{EQ17}    \partial_\alpha(  {\bf{e}}^\beta A_\beta )= 
     {\bf{e}}^\beta (\partial_{\alpha} A_\beta - \Gamma^{\mu}_{\alpha\beta} A_{\mu} )=
    {\bf{e}}^\beta (\nabla_{\alpha} A_\beta )   \end{equation}
\begin{equation} \label{EQ18}    \partial_\alpha(  {\bf{e}}_\beta A^\beta )= 
     {\bf{e}}_\beta (\partial_{\alpha} A^\beta + \Gamma^{\beta}_{\alpha\mu} A^{\mu} )=
    {\bf{e}}_\beta (\nabla_{\alpha} A^\beta )   \end{equation}
with covariant derivatives appearing naturally.
To show how the Riemann commutator also underlies the $E_3$ calculation, $\bf{\Omega}\bf{A}$ can be written
\begin{equation} \label{EQ19}
\bf{\Omega}\bf{A} = 
({\bf{e}}^\alpha \partial_{\alpha}) \times ( ( {\bf{e}}^\beta \partial_{\beta} )\times ({\bf{e}}^\gamma A_{\gamma} )) -
({\bf{e}}^\alpha \partial_{\alpha}) ( {\bf{e}}^\beta \partial_{\beta} ) {\mbox{\huge{.}}} ( {\bf{e}}_\gamma A^{\gamma}) +
({\bf{e}}^\alpha \partial_{\alpha}) {\mbox{\huge{.}}} ( {\bf{e}}^\beta \partial_{\beta}) ( {\bf{e}}_\gamma A^{\gamma} )
\end{equation}
and to extract components we just need to shift ${\bf{e}}$'s and $\partial$'s around.  Since the same shifts are 
needed for each term of (\ref{EQ19}), we just temporarily suppress all dots and crosses and consider the rank 
three tensor
\begin{equation} \label{EQ20}  ({\bf{e}}^\alpha \partial_{\alpha}) ( {\bf{e}}^\beta \partial_{\beta}) 
( {\bf{e}}_\gamma A^{\gamma} ) \end{equation}
as exemplar of each of the terms of (\ref{EQ19}). As a point on notation we adopt the principle that all regular 
partials as well as covariant operations act fully to the right regardless of any bracketting inserted purely for 
algebraic clarity.
We saw above that an ordinary partial derivative becomes a covariant derivative when pushed to the right 
past a base vector.  Considering just the last two elements of expression (\ref{EQ20}) , if one pushes the partial 
$\partial_\beta$ past ${\bf{e}}_\gamma$ a rank 2 tensor $\nabla_\beta A^\gamma$ is produced.  The
partial $\partial_{\alpha}$ still standing to the left, can now be pushed successively past each of the base 
vectors ${\bf{e}}^\beta$ and ${\bf{e}}_\gamma$ linked to this rank 2 tensor. Each 'push-past' gives a term 
with a Christoffel symbol, with finally a term involving the partial derivative of $\nabla_\beta A^\gamma$. 
The three terms comprise the three terms of the covariant derivative of $\nabla_\beta A^\gamma$. Thus 
altogether we arrive at a rank 3 tensor, namely the second covariant derivative of $\bf{A}$. Thus 
\begin{equation} \label{EQ21} 
({\bf{e}} ^\alpha \partial_{\alpha}) ( {\bf{e}} ^\beta \partial_{\beta}) ( {\bf{e}} _\gamma A^{\gamma} ) =
({\bf{e}} ^\alpha \partial_{\alpha}) ( {\bf{e}} ^\beta {\bf{e}} _\gamma \nabla_{\beta}  A^{\gamma} ) =
({\bf{e}} ^\alpha {\bf{e}} ^\beta {\bf{e}} _{\gamma} ) \nabla_{\alpha} \nabla_{\beta}  A^{\gamma}    \end{equation}
The linear combination of dot and cross vector operations represented in (\ref{EQ19}) can then be applied 
directly to the triplet of base vectors of (\ref{EQ21}), since the three base vectors are now in juxtaposition:
\begin{equation} \label{EQ22}    {\bf{e}}^\alpha \times ({\bf{e}}^\beta \times {\bf{e}}_{\gamma}) -
   {\bf{e}}^\alpha {\bf{e}}^\beta \mbox{\huge{.}}  {\bf{e}}_{\gamma} +
  {\bf{e}}^\alpha \mbox{\huge{.}} {\bf{e}}^\beta {\bf{e}} _{\gamma} =
 \delta ^{\alpha}_{\gamma} \bf{e} ^{\beta} - \delta ^{\beta}_{\gamma} \bf{e} ^{\alpha}  \end{equation}
Carrying out the replacement indicated by (\ref{EQ22}) on the vector triple of (\ref{EQ21}) one obtains
\begin{equation} \label{EQ23} {\bf{\Omega A}} = (\delta ^{\alpha}_{\gamma} {\bf{e}} ^{\beta} - \delta ^{\beta}_{\gamma} 
{\bf{e}} ^{\alpha})\nabla_{\alpha} \nabla_{\beta} A^{\gamma} =  
  - {\bf{e}}^\beta  (\nabla_{\beta} \nabla_{\gamma} - \nabla_{\gamma} \nabla_{\beta} ) A^{\gamma} = 
 - {\bf{e}}^\beta R_{\beta \gamma}  A^{\gamma}  \end{equation}
In $E_3$, which is Riemann--flat, the two covariant derivatives commute and the `identity', Equation (\ref{EQ01}),
is true regardless of whether or not, in the $E_3$, we have chosen Cartesian coordinates or orthogonal 
curvilinear coordinates, or unparticularised curvilinear coordinates.

%%%%%%%%%%%%%%%%%%%%%%%%%%%%%%%%%%%%%%%%%%%%%%%%%%%%%%%%%%%%%%%%%%%%%%%5%%%%%%%%%%%%%%%%%%%
%%%%%%%%%%%%%%%%%%%%%%%%%%%%%%%%%%%%%%%%%%%%%%%%%%%%%%%%%%%%%%%%%%%%%%%%%%%%%%%%%%%%%%%%%%%
(c): \ \ \ {\bf{A Euclidean space addendum.}} \ \ \ 
%%%%%%%%%%%%%%%%%%%%%%%%%%%%%%%%%%%%%%%%%%%%%%%%%%%%%%%%%%%%%%%%%%%%%%%%%%%%%%%%%%%%%%%%%%%
If one wishes to see explicitly appear 
the actual terms of the Ricci tensor in terms of Christoffel symbols, this most easily 
follows from the following minimal choice of shifts of partials on the first 
expression occurring in (\ref{EQ21}). The first partial is shifted to the left and the second partial 
to the right, using (\ref{EQ15}) and (\ref{EQ16}), to produce:
\begin{equation} \label{EQ24}  (\partial_{\alpha} + \Gamma^{\mu}_{\alpha \mu})\ {\bf{e}^\alpha \bf{e}^\beta \bf{e}_{\rho}}\ 
(\delta^{\rho}_{\gamma}\partial_{\beta} + \Gamma^{\rho}_{\beta \gamma}) A^{\gamma}  \end{equation}
To produce the terms of  $\bf{\Omega}\bf{A}$ the embedded triple of base vectors appearing in (\ref{EQ24}) is
replaced by the expression (\ref{EQ22}) giving directly
\begin{equation} \label{EQ25}   {\bf{e}}^\beta ({\bf{\Omega A}})_{\beta}
 = {\bf{e}} ^{\beta} ( \Gamma^{\mu}_{\rho\mu} \Gamma^{\rho}_{\beta\gamma}
 - \Gamma^{\mu}_{\rho\beta} \Gamma^{\beta}_{\mu\gamma} +\Gamma^{\rho}_{\beta \gamma ,\rho}
 - \Gamma^{\alpha}_{\alpha\gamma,\beta}) A^{\gamma}
 = -{\bf{e}}^\beta R_{\beta \gamma}  A^{\gamma}
\end{equation}
regardless of what curvilinear coordinate system is chosen and whether or not that choice is an orthogonal 
one. Note that the remaining base vector $ \bf{e}^\beta$ must be systematically moved to the left in each term 
in the intermediate expressions. Thus in $E_3$ which is Riemann and Ricci flat the usual physics identity is 
established with any choice of curvilinear coordinates.

%%%%%%%%%%%%%%%%%%%%%%%%%%%%%%%%%%%%%%%%%%%%%%%%%%%%%%%%%%%%%%%%%%%%%%%%%%%%%%%%%%%%%%%%%%%%%%%%%%5
%%%%%%%%%%%%%%%%%%%%%%%%%%%%%%%%%%%%%%%%%%%%%%%%%%%%%%%%%%%%%%%%%%%%%%%%%%%%%%%%%%%%%%%%%%%%%%%%%%%
(d): \ \ \ {\bf{Applications.}} \ \ \   
%%%%%%%%%%%%%%%%%%%%%%%%%%%%%%%%%%%%%%%%%%%%%%%%%%%%%%%%%%%%%%%%%%%%%%%%%%%%%%%%%%%%%%%%%%%%%%%%%%%
Our results do have an immediate application, for example, in producing 
a simpler tensor result for the Laplacian of a vector field in an arbitrary curvilinear coordinate 
system. After obtaining by direct calculation

\begin{equation} \label{EQ26}    (\nabla^{2} {\bf{A}})^{\mu}  =   \nabla^{2}(A^{\mu}) + 2 g^{\alpha\beta} 
\Gamma^{\mu}_{\beta\lambda} \partial_{\alpha}A^{\lambda} + g^{\alpha\beta} ( \Gamma^{\mu}_{\alpha\tau} 
 \Gamma^{\tau}_{\beta\lambda} - \Gamma^{\mu}_{\tau\lambda} \Gamma^{\tau}_{\alpha\beta} 
+\Gamma^{\mu}_{\beta \lambda ,\alpha})A^{\lambda}
\end{equation}
one recognizes the bracketed factor on the right as comprising most of the Riemann tensor.  Consequently 
in $E_3$, after setting the Riemann tensor to zero, expression (\ref{EQ26}) can be rewritten as
\begin{equation} \label{EQ27}  (\nabla^{2} A)^{\mu}  =   \nabla^{2}(A^{\mu})
+ 2 g^{\alpha\beta} \Gamma^{\mu}_{\beta\lambda} \partial_{\alpha}A^{\lambda} 
 + g^{\alpha\beta} \Gamma^{\mu}_{\beta \alpha ,\lambda} A^{\lambda} \end{equation}
a rather simpler result than (\ref{EQ26}).  In each of  (\ref{EQ26}) and (\ref{EQ27}) the first term on the 
right is the well--known formal expression
\begin{equation} \label{EQ028} \nabla^{2}(A^{\mu})  = (\partial^{\sigma} - g^{\kappa \pi} \Gamma^{\sigma}_{\kappa\pi} ) 
\partial_{\sigma} A^{\mu} = g^{-1/2}\ \partial_{\sigma} g^{1/2}\ \partial^{\sigma} A^{\mu}
\end{equation}
The form of this term is exactly that for the Laplace-Beltrami operator on a scalar; thus the extra terms present on
the RHS of (\ref{EQ27}) represent the extension of the Laplace-Beltrami operator to a contravariant vector field. 
If we apply (\ref{EQ27}) using say spherical polars we obtain the 
standard result for the Laplacian of a vector field as given for example in Morse and 
Feshbach (see p.116 of {\cite{bib-name1}}). (In making such comparisons, we must recall 
that, almost always in physics texts using orthogonal curvilinear 
coordinate systems, all vector 
expressions are related to a triple of {\bf unit} base vectors so that vector components with respect to the 
unit--set will in general differ from both proper contravariant components and proper covariant components.)

%%%%%%%%%%%%%%%%%%%%%%%%%%%%%%%%%%%%%%%%%%%%%%%%%%%%%%%%%%%%%%%%%%%%%%%%%%%%%%%%%%%%%%%%%%%%%%%%%%%%%%%%%55
%%%%%%%%%%%%%%%%%%%%%%%%%%%%%%%%%%%%%%%%%%%%%%%%%%%%%%%%%%%%%%%%%%%%%%%%%%%%%%%%%%%%%%%%%%%%%%%%%%%%%%%%%%%
\ {\bf{Conclusion.}} \ \ 
%%%%%%%%%%%%%%%%%%%%%%%%%%%%%%%%%%%%%%%%%%%%%%%%%%%%%%%%%%%%%%%%%%%%%%%%%%%%%%%%%%%%%%%%%%%%%%%%%%%%%%%%%%%
The identity $\bf{\Omega}\bf{A}$ = 0 is true without any restriction on coordinate system used, but only 
in a Euclidean $E_3$.  In a Riemannian $V_3$, it becomes
\begin{equation} \label{EQ29} ( \mbox{curl (curl}\ {\bf{A})})^{\lambda} - \nabla^{\lambda}\nabla_{\mu} A^{\mu} + 
\nabla_{\mu}\nabla^{\mu} A^{\lambda} = -A^{\rho} R_{\rho}^{\lambda}  \end{equation}
with the rank--2 tensor $R$ being the Ricci tensor.
%%

%% ==================================================================
%%    ACKNOWLEDGEMENT
%%
%%\acknowledgments{whatever--whatsoever}
%%
\ \ \ {\bf{Acknowledgment.}}\ \ The author wishes to express his gratitude to the chair and faculty of the 
University of Otago, Physics Department for their kind hospitality, and especially to Dr T.C.A.Molteno and 
his research group.
%%
%%
%%

%%
%% ==================================================================
%%
%%    BIBLIOGRAPHY
%%
%%\bibliography{}

\begin{thebibliography} {99}
\bibitem{bib-name3} There seems to be no agreed convention for an overall algebraic sign in the definition of the 
Riemann tensor.
\bibitem{bib-name6} It seems worth noting that both the Riemann and Ricci tensors occur in the Weitzenboeck identity 
evaluating the difference of two independent definitions of the Laplacian in the language of differential 
forms. Besides what one might call the usual Laplacian, there exist also the Bochner Laplacian, the Hodge 
Laplacian, the Lichnerowicz Laplacian, the Conformal Laplacian. The aims and results of this paper 
seem far removed from the differential geometry discussions, and any attempt in this paper to make 
connections between our work and such discussions would be at odds with our intent of elucidating 
the truth or untruth of the identity (\ref{EQ01}), as part of commonly used vector and tensor analysis. In such a 
context the Laplacian operator must be the contracted double covariant derivative operator
\begin{equation}\label{EQ30}
\nabla^{\mu}\nabla_{\mu} =  g^{\mu \nu} \nabla_{\mu}\nabla_{\nu} =   \nabla_{\mu}\nabla^{\mu}
\end{equation}
and is clearly a well--defined invariant operator.
\end{thebibliography}
%%\bibliographystyle{apsrev4-1}
%\begin{equation} \label{EQ03} B^\alpha = \varepsilon^{\alpha\beta\gamma}\ \nabla_\beta\ ( (g^{-1/2}) 
% A _\gamma)
%\end{equation}

\end{document}